\documentstyle{article}
\begin{document}
\title{Perturbations in inflationary cosmologies 
with smooth exit}
\author{Daniel Tilley and Roy Maartens
\\
\\
{ \footnotesize School of Computer Science and Mathematics, 
University of Portsmouth, Portsmouth PO1 2EG, Britain}}
\maketitle

\begin{abstract}
We give a comprehensive analysis of how scalar and tensor perturbations evolve in cosmologies with a
smooth transition from power-law-like and de Sitter-like inflation to a 
radiation era.   
Analytic forms for the super-horizon and sub-horizon perturbations
in the inflationary and radiation dominated eras are
found. 
\end{abstract}
%\vfill
\renewcommand{\thesection}{\Roman{section}}
\renewcommand{\thesubsection}{\Alph{subsection}}
%\newpage
\section{Introduction}
In \cite{M11}, the evolution of perturbations for a decaying vacuum 
cosmology given in \cite{M9}, with smooth exit from inflation to 
radiation-domination, was found exactly on 
super-horizon scales. The solution confirmed the standard result 
\cite{dm,M8,M1,klop2,M12}
within
gauge invariant perturbation theory that super-horizon 
density perturbations are  
strongly amplified through the transition from inflation to radiation.
This transition is 
often approximated as an instantaneous jump 
\cite{dm,M8,M12}, since super-horizon modes change on a  
timescale that is much greater than the transition time. To avoid problems 
in the matching conditions for a jump transition, we consider only 
smooth transitions.

Given the variety of methods and models that have been used, we believe 
it is useful to do a comparative study, using the same method on two
broad classes of transition models, i.e. those based on power-law-like
and de Sitter-like inflation. 
We use the class of
power-law-like models first 
given  
in \cite{M2}. The second class of models is a generalisation of the 
models of \cite{M1,M2} with a de Sitter-like inflationary era. 
The models, together with the neccessary equations in perturbation theory, 
are discussed in section 2.
In section 3 we present our results for scalar and tensor perturbations 
on super-horizon scales, and show how they confirm the standard 
predictions. 
In section 4 we present solutions for scalar and tensor 
perturbations on sub-horizon scales. However,            
toy models that evolve smoothly from inflation to radiation dominated 
eras are only applicable 
for super-horizon modes around the time of transition, since on 
sub-horizon scales, the dynamics of the reheating  
era have a significant 
effect. The solutions for sub-horizon scales therefore do not apply 
during the transition era.
Section 5 gives some concluding remarks. 

Throughout 
this paper we use units in which $8\pi G=1=c$.

\section{The models and perturbation equations}
A flat Friedmann-Lemaitre-Robertson-Walker (FLRW) background has metric
\begin{equation}
ds^{2}=-dt^{2}+a^{2}(t)[dx^{2}+dy^{2}+dz^{2}]\,,\label{1}
\end{equation}
where $t$ is the physical time and $a(t)$ is the expansion scale factor.
The background pressure $p$, energy density $\rho$ and  
Hubble parameter $H=\dot{a}/a$ (where a dot denotes a derivative 
with respect to cosmic time $t$)
satisfy the field equations
\begin{eqnarray}
p&=&-2\dot{H}-3H^{2}\,,\label{3} \\
\rho&=&3H^{2}\,.\label{4}
\end{eqnarray}
The pressure index and effective adiabatic sound speed are defined by
\begin{eqnarray}
w &=& {p\over \rho}\,,\label{a} \\
c_{\rm s}^{2} &=& {dp\over d\rho}={\dot{p}\over \dot{\rho}}
\,\label{b}\end{eqnarray}
Inflation is characterised by $\ddot{a}>0$, so that the epoch $t_{\rm e}$ 
of exit is given by $\ddot{a}(t_{\rm e})=0$. The radiation dominated  
era is characterised by $p\approx {1\over 3}\rho$, $H\approx {1/2t}$. 

Scalar and tensor perturbations lead to a metric of the form
\begin{equation}
ds^{2}=a^{2}(\eta)\{-(1+2\Phi)d\eta^{2}+[(1-2\Psi)\gamma_{ij}
+h_{ij}]dx^{i}dx^{j}\}
\,,\label{non1}\end{equation}
in the longitudinal gauge, where $\eta$ is conformal time ($d\eta=dt/a$) 
\cite{M8}. Here $\Phi$ and $\Psi$ are 
gauge-invariant
amplitudes of the scalar metric perturbations. The field equations require  
$\Phi=\Psi$ when the spatial part of the energy-momentum tensor is diagonal.
The tensor perturbations are gauge-invariantly described by 
$h_{ij}=hQ_{ij}$, where $Q_{ij}$ is a symmetric traceless three 
tensor and $h$ 
represents the amplitude of a gravitational wave.

The evolution of adiabatic scalar perturbations $\Phi$ is given 
by \cite{M8}
\begin{equation}
{d^{2}\Phi\over d\eta^{2}}+3aH(1+c_{\rm s}^{2}){d\Phi\over d\eta}-\nabla^{2}\Phi
+3a^{2}H^{2}(c_{s}^{2}-w)
\Phi=0\,.\label{c}\end{equation}
Decomposing $\Phi(\eta,\vec{x})$ into eigenmodes $\tilde{\Phi}(\eta,\vec{k})$
of the comoving Laplacian,
and then using the cosmic time $t$ and scale factor $a$ as the dynamical  
variables respectively, gives
\begin{eqnarray} &&
{d^{2}\tilde{\Phi}\over dt^{2}}+3H\left({4\over 3}+c_{\rm s}^{2}\right)
{d\tilde{\Phi}\over dt}
+\left[3H^{2}(c_{s}^{2}-w)+\left({k\over a}\right)^{2}\right]
\tilde{\Phi}=0\,,\label{c1}\\ &&
{d^{2}\tilde{\Phi}\over da^{2}}+{1\over a}\left({7\over 2}+3c_{s}^{2}-
{3\over 2}w\right)
{d\tilde{\Phi}\over da}
+\left[{3\over a^{2}}(c_{s}^{2}-w)+{k^{2}\over a^{4}H^{2}}\right]
\tilde{\Phi}=0\,,\label{c1d}\end{eqnarray}
where $k$ is the comoving wave number.
The decomposition allows us to follow the evolution of a single mode 
through the inflationary and radiation dominated eras. 
The physical length scale $\lambda$ correponding to the comoving 
wavenumber $k$ is $\lambda=2\pi a/k$. This will equal the 
Hubble radius $H^{-1}$ when 
\begin{equation}
k=2\pi aH\,.\label{khopda1}\end{equation} 
Super-horizon modes are characterised by $k\ll aH$. For these modes one 
can neglect the $k^{2}$ term in equations (\ref{c1}) and (\ref{c1d})
(and subsequent equations). The solutions of the truncated equations 
will be a good approximation to the limit of the solutions of the full 
equations, provided \cite{M19}
\begin{equation}
k^{2}\Phi\rightarrow 0\,\mbox{\rm as}~~k\rightarrow 0\,.\label{grin}\end{equation}
For adiabatic super-horizon 
scalar perturbations, the growing modes have 
a conserved quantity \cite{zop1} which can be used to determine 
$\tilde{\Phi}$: 
\begin{equation}
\zeta\equiv \tilde{\Phi}+{2\over 3(1+w)}\left(\tilde{\Phi}+{1\over H}
{d\tilde{\Phi}\over dt}\right)\Rightarrow \tilde{\Phi}=-\zeta
{H\over a}\int{{a\dot{H}\over H^{2}}}dt
\,.\label{16}\end{equation}
Equation (\ref{16}) can be used to express the 
growing perturbations at late times in terms of their early time forms 
\cite{M8}. In \cite{M1,klop2,M12}, $\zeta$ is used to estimate the 
amplification of perturbations through a smooth transition. This 
amplification has been disputed in \cite{M5} (see also \cite{kosh1,Kerr}).

The density perturbations in the longitudinal gauge are given by \cite{M8}:
\begin{equation}
{\delta\rho\over \rho}=-2\left({k^{2}\over 3a^{2}H^{2}}+1\right)\tilde{\Phi}
-{2\over H}{d\tilde{\Phi}\over dt}
\,.\label{d}\end{equation}
The Fourier mode $\tilde{h}$ of the amplitude $h$ of the tensor perturbations 
satisfies the 
equation \cite{M8}
\begin{equation}
{d^{2}\tilde{h}\over dt^{2}}+3H{d\tilde{h}\over dt}+
\left({k\over a}\right)^{2}\tilde{h}=0\,,\label{e}\end{equation}
which may also be re-written with $a$ as the dynamical variable:
\begin{equation}
a^{2}{d^{2}\tilde{h}\over da^{2}}+\left(4a+{a^{2}\over H}{dH\over da}\right)
{d\tilde{h}\over da}+
{k^{2}\over a^{4}H^{2}}\tilde{h}=0\,.\label{e1d}\end{equation}
For super-horizon modes one can reduce to quadrature:
\begin{equation}
\tilde{h}=C+D\int{dt\over a^{3}}=C+D\int{da\over a^{4}H}\,,\label{bake45}\end{equation}
where $C$ and $D$ are constant. It follows that $D$ corresponds to the
decaying mode and therefore there is no amplification in
long-wavelength tensor perturbations \cite{dm,M8}.

\subsection{Power-law-like models}
In \cite{M2} a simple class of power-law-like inflation models, which 
exit smoothly to a radiation era, is given by 
\begin{equation}
a(t)={a_{\rm e}\over \sqrt{t_{\rm e}}}(1+m)^{(2n-1)/2}t^{n}(t+mt_{\rm e})^{(1-2n)/2}
\,,\label{2}\end{equation}
where the parameter $n$ satisfies $n>1$ to acheive inflationary expansion, and  
$$m={n+\sqrt{n(2n-1)}\over 2n(n-1)}\,.$$ 
For $t\ll mt_{\rm e}$, 
$$a\approx a_{\rm e}\left(1+{1\over m}\right)^{(2n-1)/2}\left({t\over t_{\rm e}}\right)^{n}\,,$$ 
showing the early power-law-like inflation era. 
For $t\gg mt_{\rm e}$,  
$$a\approx a_{\rm e}\left(1+m\right)^{(2n-1)/2}\left({t\over t_{\rm e}}\right)^{1/2}\,,$$ 
as in a radiation dominated era.
Thus we have a 
cosmology which evolves smoothly from power-law-like inflation
to a radiation dominated era.
The Hubble rate follows on differentiating equation 
(\ref{2}):
\begin{equation}
H(t)={t+2nmt_{\rm e}\over 2t(t+mt_{\rm e})}\,.\label{2a}
\end{equation}
Using equations (\ref{3}) and (\ref{4}), equation (\ref{2a}) gives
\begin{eqnarray}
p &=& {t^{2}+4nmt_{\rm e}t-4n(3n-2)m^{2}
t_{\rm e}^{2}\over 4t^{2}(t+mt_{\rm e})^{2}}
\,,\label{5} \\
\rho &=& {3(t+2nmt_{\rm e})^{2}\over 4t^{2}(t+mt_{\rm e})^{2}}\,.\label{6}
\end{eqnarray}
It follows that the pressure index and effective adiabatic sound speed
given by equations (\ref{a}) and (\ref{b}) are
\begin{eqnarray}
w &=& {t^{2}+4nmt_{\rm e}t-4n(3n-2)m^{2}t_{\rm e}^{2}\over
3t^{2}+12nmt_{\rm e}t+12n^{2}m^{2}t_{\rm e}{}^{2}}\,,\label{7} \\
c_{\rm s}^{2} 
&=&
{t(t+mt_{\rm e})(t+2nmt_{\rm e})-
[t^{2}+4nmt_{\rm e}t-4n(3n-2)m^{2}t_{\rm e}^{2}]
(2t+mt_{\rm e})
\over 3t(t+mt_{\rm e})(t+2nmt_{\rm e})-3(t+2nmt_{\rm e})^{2}(2t+mt_{\rm e})}\,.\label{8}
\end{eqnarray}
 
The evolution of scalar and tensor perturbations for 
this model is given by equations (\ref{c1}), (\ref{d}) and (\ref{e}). 
Using equations
(\ref{2}), (\ref{2a}), 
(\ref{7}) and (\ref{8}), we get
\begin{equation} 
t^{2}{d^{2}{\tilde{\Phi}}\over dt^{2}} 
+t\left[{t+2nmt_{\rm e}\over 2(t+mt_{\rm e})}\right]
U(t){d{\tilde{\Phi}}\over dt}+
\left[{t+2nmt_{\rm e}\over 2(t+mt_{\rm e})}\right]^{2}
V(t)\tilde{\Phi}
=0\,,\label{12e}\end{equation}
where
\begin{eqnarray}
U(t)& = &
{5t(t+mt_{\rm e})(t+2nmt_{\rm e})-(5t^{2}+20nmt_{\rm e}t+4n(n+2)m^{2}t_{\rm e}^{2})
(2t+mt_{\rm e})\over t(t+mt_{\rm e})(t+2nmt_{\rm e})
-(t+2nmt_{\rm e})^{2}(2t+mt_{\rm e})}
\,,\nonumber\\
V(t) & = & 
{t(t+mt_{\rm e})(t+2nmt_{\rm e})-[t^{2}+4nmt_{\rm e}t
-4n(3n-2)m^{2}t_{\rm e}^{2}]
(2t+mt_{\rm e})\over 
t(t+mt_{\rm e})(t+2nmt_{\rm e})-(t+2nmt_{\rm e})^{2}(2t+mt_{\rm e})}
\nonumber\\ &&{}- 
{t^{2}+4nmt_{\rm e}t-4n(3n-2)m^{2}t_{\rm e}^{2}\over
(t+2nmt_{\rm e})^{2}}+
k^{2}
{t_{\rm e}(1+m)^{1-2n}t^{2-2n}(t+mt_{\rm e})^{2n+1}
\over a_{\rm e}^{2}(t+2nmt_{\rm e})^{2}}\,,\nonumber\end{eqnarray}
and
\begin{equation}
{\delta{\rho}\over \rho}=-2\left[
{4k^{2}t_{\rm e}(1+m)^{1-2n}t^{2-2n}(t+mt_{\rm e})^{2n+1}
\over 3a_{\rm e}^{2}(t+2nmt_{\rm e})^{2}}+1\right]\tilde{\Phi}
-\left[{4t(t+mt_{\rm e})\over t+2nmt_{\rm e}}\right]{d{\tilde{\Phi}}\over dt}\,,
\label{12a}\end{equation}
and
\begin{eqnarray} &&
t^{2}{d^{2}\tilde{h}\over dt^{2}}+
3t\left[{t+2nmt_{\rm e}\over 2(t+mt_{\rm e})}\right]
{d\tilde{h}\over dt}\nonumber\\&&{}
+\left[{k^{2}t_{\rm e}(1+m)^{1-2n}(t+mt_{\rm e})^{2n-1}t^{2-2n}
\over a_{\rm e}^{2}}\right]\tilde{h}=0\,.\label{lm1}\end{eqnarray}
 
\subsection{de Sitter-like models}
In \cite{M2} a new exact 
inflationary solution is given, using $a$ as the 
effective dynamical variable. The Hubble rate is given by
\begin{equation}
H(a)={2H_{\rm e}a_{\rm e}^{2}\over a^{2}+a_{\rm e}^{2}}\,,\label{2b1}
\end{equation}
where $H_{\rm e}$ is the Hubble rate at exit. 
A similar model [but one which does not 
produce an exact solution for super-horizon scalar perturbations, 
unlike (\ref{2b1}) \cite{M11}) is proposed in \cite{M1}:
\begin{equation}
H(a)={2^{1/2}H_{\rm e}a_{\rm e}^{2}\over (a^{4}+a_{\rm e}^{4})^{1/2}}\,.\label{2a1}
\end{equation}
For 
both Hubble rates,
when $a\ll a_{\rm e}$, $H$ is approximately constant as in a de Sitter 
inflation 
era, and for $a\gg a_{\rm e}$, the Hubble rate decays like 
$a^{-2}$, as in a radiation dominated era. Thus for both models 
we have a 
cosmology which evolves smoothly from inflation to 
a radiation dominated era.

Here we generalise these models by considering the class of Hubble rates
\begin{equation}
H(a)={2^{1/\ell}a_{\rm e}^{2}H_{\rm e}\over (a^{2\ell}+a_{\rm e}^{2\ell})^{1/\ell}}
\,,\label{2c1}\end{equation}
parametrised by $\ell$ ($\ell>0$). For equation (\ref{2b1}), $\ell=1$, 
and for 
(\ref{2a1}), $\ell=2$. 
Using equation (\ref{2c1}), $H\approx$ constant for $a\ll a_{\rm e}$ 
(inflation), and $H\propto a^{-2}$ for $a\gg a_{\rm e}$ 
(radiation domination), confirming that the model has the desired properties 
for all positive $\ell$. 

Using equations (\ref{3}) and (\ref{4}), equation (\ref{2c1}) gives
\begin{eqnarray}
p &=& {{2^{2/\ell}H_{\rm e}^{2}a_{\rm e}^{4}(a^{2\ell}
-3a_{\rm e}{}^{2\ell}})\over 
(a^{2\ell}+a_{\rm e}^{2\ell}
)^{1+2/\ell}}\,,\label{5a1} \\
\rho &=& {2^{2/\ell}3H_{\rm e}^{2}a_{\rm e}^{4}\over 
(a^{2\ell}+a_{\rm e}^{2\ell})^{2/\ell}}\,.\label{6a1}
\end{eqnarray}
The pressure index and effective adiabatic sound speed  
given by equations (\ref{a}) and (\ref{b}) are
\begin{eqnarray}
w &=& {1\over 3}\left({a^{2\ell}-3a_{\rm e}^{2\ell}\over 
a^{2\ell}+a_{\rm e}^{2\ell}}\right)\,,\label{7a1} \\
c_{\rm s}^{2} &=& 
{1\over 3}\left[{a^{2\ell}-(2\ell+3)a_{\rm e}^{2\ell}\over a^{2\ell}+
a_{\rm e}^{2\ell}}\right]\,.\label{8a1}
\end{eqnarray}
Applying equation 
(\ref{khopda1}) to equation (\ref{2c1}) and taking 
$k_{\rm e}=2\pi a_{\rm e}H_{\rm e}$ 
(the 
comoving wavenumber of the Hubble radius at exit), the epoch $a_{-}$ 
of leaving and $a_{+}$ of re-entering 
the Hubble radius are
given exactly in this class of models by
\begin{equation}
{a_{\pm}\over a_{\rm e}}={k_{\rm e}\over k}\left[1\pm\sqrt{1-
\left({k\over k_{\rm e}}\right)^{2\ell}}\right]^{1/\ell}\,.\label{doppy}
\end{equation}
Scales with $k\geq k_{\rm e}$ never cross the Hubble radius and remain 
sub-horizon. 
For super-horizon scales with
$k\ll k_{\rm e}$, equation (\ref{doppy}) reduces to
$${a_{\pm}\over a_{\rm e}}\approx\left({2^{1/\ell}k_{\rm e}\over k}\right)^{\pm 1}\,.$$

The scalar and tensor perturbations for this model are governed  by  
equations (\ref{c1d}), (\ref{d}) and (\ref{e1d}). 
Using equations (\ref{2c1}), (\ref{7a1}) and 
(\ref{8a1}) we get
\begin{eqnarray}&&
a^{2}{d^{2}\tilde{\Phi}\over da^{2}}+a\left[{4a^{2\ell}-(2\ell-2)a_{\rm e}^{2\ell}
\over a^{2\ell}+a_{\rm e}^{2\ell}}\right]
{d\tilde{\Phi}\over da}\nonumber\\
&&+\left[
\pi^{2}\left({k\over k_{\rm e}}\right)^{2}{(a^{2\ell}
+a_{\rm e}^{2\ell})^{2/\ell}\over 
2^{2/\ell-2}
a_{\rm e}^{2}a^{2}}
-{2\ell a_{\rm e}^{2\ell}\over (a^{2\ell}+a_{\rm e}^{2\ell})}\right]
\tilde{\Phi}=0
\,,\label{20}\end{eqnarray}
and
\begin{equation}
{\delta{\rho}\over \rho}=-2\left[{1\over 3}\left({k
\over k_{\rm e}}\right)^{2}{(a^{2\ell}+a_{\rm e}^{2\ell})^{2/\ell}\over 
2^{2/\ell}a_{\rm e}^{2}a^{2}}+1\right]\tilde{\Phi}-{2a{d\tilde{\Phi}\over da}}
\,,\label{12a1}\end{equation}
and
\begin{equation}
a^{2}{d^{2}{\tilde{h}}\over da^{2}}+2a\left[{a^{2\ell}+
2a_{\rm e}^{2\ell}\over 
a^{2\ell}+a_{\rm e}^{2\ell}}\right]
{d{\tilde{h}}\over da}+\left[\pi^{2}\left({k\over k_{\rm e}}\right)^{2}
{(a^{2\ell}+a_{\rm e}^{2\ell})^{2/\ell}\over 
2^{2/\ell-2}a_{\rm e}^{2}a^{2}}\right]\tilde{h}=0\,.\label{23}\end{equation}

\section{Super-horizon perturbations}
For the modes which leave the Hubble radius (during inflation),
while they remain outside the Hubble radius, we can neglect the $k$-term in 
equations (\ref{12e}), (\ref{12a}), (\ref{lm1}), (\ref{20}), (\ref{12a1}) 
and (\ref{23}). 
Only equations (\ref{lm1}) and (\ref{23}), using (\ref{bake45}), 
give exact solutions for the truncated equations.
However, using \cite{M4,M10} we can 
give analytic forms for the solutions in the inflationary and 
radiation dominated eras for equations 
(\ref{12e}), (\ref{12a}), 
(\ref{20}) and (\ref{12a1}),
since they reduce to Bessel forms.   
For all solutions, $A_{j}$, $B_{j}$, $j=1,2,...$ 
are used to denote arbitrary constants.
For equation (\ref{grin}) to be satisfied it is required that for
scalar perturbations $A_{j}$, $B_{j}\sim k^{q}$ as $k\rightarrow 0$,
where $q>-2$. For all scalar perturbation solutions this is assumed 
to be the case. 
\subsection{Power-law-like model}
For $t\ll mt_{\rm e}$ equation (\ref{12e}) has solution
\begin{equation}
\tilde\Phi\approx A_{1}+B_{1}\left({t_{\rm e}\over t}\right)^{n+1}
\,,\label{mz1}\end{equation}
describing modes that leave well before exit,
while for $t\gg mt_{\rm e}$
\begin{equation}
\tilde{\Phi}\approx A_{2}+B_{2}\left({t_{\rm e}\over t}\right)^{3/2}
\,,\label{mz2}\end{equation}
describing modes that remain super-horizon in the radiation era.
In equations (\ref{mz1}) and (\ref{mz2}), $A_{j}$ correspond to 
the constant modes, and $B_{j}$ to the 
decaying modes.

The conserved quantity given by equation (\ref{16}), 
applied to 
the case of the model 
given by (\ref{2}), leads to
\begin{equation}
{\tilde{\Phi}}
=\zeta{(t+mt_{\rm e})^{(2n-3)/2}(t+2nmt_{\rm e})\over t^{n+1}}
\int{{t^{n}(t^{2}+4nmt_{\rm e}t+2nm^{2}t_{\rm e}{}^{2})
\over (t+2nmt_{\rm e})^{2}(t+mt_{\rm e})^{(2n-1)/2}}}dt
\,,\label{1H1}\end{equation}
where we neglect the decaying contributions (and therefore no
constant of integration arises from the integral).
Amplification 
of the 
scalar growing modes according to equations (\ref{mz1}) and (\ref{mz2})
is given by the factor $\alpha=A_{2}/A_{1}$. Using equation (\ref{1H1}),
we find that
$$A_{1}={\zeta\over (n+1)}\,,$$
and 
$$A_{2}= {2\over 3}\zeta\,.$$
Therefore 
\begin{equation}
\alpha ={2\over 3}(n+1)>{4\over 3}\,,\label{kruno}\end{equation}
so that amplification is only significant for large $n$. This is in 
agreement with results based on instantaneous transition models 
\cite{dm,M8,M12}.

The density perturbations in the inflationary 
and 
radiation dominated eras, are found analytically, using equations 
(\ref{12a}), (\ref{mz1}) and (\ref{mz2}).
For modes that leave the Hubble radius well before exit ($t\ll mt_{\rm e}$)
\begin{equation}
{\delta{\rho}\over \rho}\approx -2A_{1}+{2\over n}B_{1}
\left({t_{\rm e}\over t}\right)^{n+1}\,,\label{mz3}\end{equation}
while for modes that are still super-horizon during radiation 
domination ($t\gg mt_{\rm e}$)
\begin{equation}
{\delta{\rho}\over \rho}\approx -2A_{2}+4B_{2}\left({t_{\rm e}\over t}
\right)^{3/2}\,.\label{mz4}\end{equation}
The density perturbations undergo the same amplification (\ref{kruno}) as 
the potential perturbations.

The tensor perturbations are given by equation (\ref{bake45}):
\begin{equation}
\tilde{h}=A_{3}+B_{3}\int{t^{-3n}(t+mt_{\rm e})^{3(2n-1)/2}}dt
\,.\label{15c}\end{equation}
During power-law inflation ($t\ll mt_{\rm e}$) 
\begin{equation}
\tilde{h}\approx A_{3}+B_{3}{m^{(6n-3)/2}t_{e}^{-1/2}\over 1-3n}
\left({t_{\rm e}\over t}\right)^{3n-1}
\,,\label{mz5}\end{equation}
while for those modes that remain super-horizon during radiation 
domination ($t\gg mt_{\rm e}$)
\begin{equation}
\tilde{h}\approx A_{3}-2B_{3}t^{-1/2}\,.\label{mz6}
\end{equation}
The results (\ref{15c})-(\ref{mz6}) imply that tensor perturbations
do not grow on super-horizon scales.

\subsection{de Sitter-like model}
For modes that leave the Hubble radius when $a\ll a_{\rm e}$, 
equation (\ref{20}) has 
super-horizon solutions
\begin{equation}
\tilde\Phi\approx A_{4}\left({a\over a_{\rm e}}\right)^{2\ell}
+B_{4}{a_{\rm e}\over a}
\,.\label{zo1}\end{equation}
For super-horizon modes during radiation domination ($a\gg a_{\rm e}$) 
\begin{equation}\tilde{\Phi}\approx \left({a_{\rm e}\over a}\right)^{3/2}
Z_{-3/2\ell}\left[-i\sqrt{2\over \ell}
\left({a_{\rm e}\over a}\right)^{\ell}\right]\,,\label{zo2a}\end{equation}
where $Z_{\nu}$ denotes a linear combination
of the Bessel functions $J_{\nu}$ and $Y_{\nu}$.
Using the asymptotic forms of the Bessel functions 
\cite{M4,M10}, equation (\ref{zo2a}) reduces to
\begin{equation}
\tilde{\Phi}\approx A_{5}+B_{5}\left({a_{\rm e}\over a}\right)^{3}\,.\
\label{zo2}\end{equation}
It follows that in the inflationary era, $|\tilde{\Phi}|$ grows as 
$a^{2\ell}$, while in the radiation dominated era, 
$|\tilde{\Phi}|$ is approximately constant (while the scales are still 
super-horizon). Thus growing super-horizon scalar perturbations are 
strongly amplified 
during inflation and then remain approximately constant after inflation. 
The amplification factor is clearly much greater than that for the 
power-law-like model.

We can calculate the amplification using equation (\ref{16}), which gives 
\begin{equation}
\tilde{\Phi}={2\zeta\over a(a^{2\ell}+a_{\rm e}^{2\ell})^{1/\ell}}
\int{a^{2\ell}(a^{2\ell}+a_{\rm e}^{2\ell})^{1/\ell}\over 
(a^{2\ell}+a_{\rm e}^{2\ell})}
da\,.\label{18}\end{equation}
Evaluating the integral in the inflationary and radiation era's and 
using equations 
(\ref{zo1}) and (\ref{zo2}), we find 
$$A_{4}={2\zeta\over 2\ell +1}\,,$$
and
$$A_{5}={2\zeta\over 3}\,.$$
The amplification from the epoch $a_{\rm i}$ ($a_{\rm i}\ll a_{\rm e}$) 
to the
radiation era is thus
\begin{equation}
\alpha={1\over 3}(2\ell+1)\left({a_{\rm e}\over a_{\rm i}}\right)^{2\ell}
\,.\label{kruno2}\end{equation}
The explicit dependence in (\ref{kruno2}) of the amplification on the 
initial epoch $a_{\rm i}$ shows that amplification, and thus the spectrum, 
is not scale invariant.

The density perturbations are given 
using equations (\ref{12a1}), (\ref{zo1}) and  
(\ref{zo2a}).
For modes beyond the Hubble radius during inflation
\begin{equation}
{\delta{\rho}\over \rho}\approx -2(2\ell+1)A_{4}\left({a\over a_{\rm e}}
\right)^{2\ell}\,,\label{zo3}\end{equation}
while for super-horizon modes during the radiation dominated era ($a\gg a_{\rm e}$)
\begin{eqnarray}
{\delta{\rho}\over \rho}& \approx & 2i\sqrt{2\ell}\left({a_{\rm e}\over a}
\right)^{\ell +3/2}Z_{(2\ell-3)/2\ell}\left[-i
\sqrt{2\over \ell}\left({a_{\rm e}\over a}\right)^{\ell}\right]
\nonumber\\&&{}-2\left({a_{\rm e}\over a}\right)^{3/2}
Z_{-3/2\ell}\left[-i\sqrt{2\over \ell}\left({a_{\rm e}\over a}\right)^{\ell}
\right]\,,\end{eqnarray}
which leads to 
\begin{equation}
{\delta{\rho}\over \rho}\approx -2A_{5}
+4B_{5}\left({a_{\rm e}{}\over a}
\right)^{3}\,.\label{zo4}\end{equation}
During inflation ($a\ll a_{\rm e}$), $\delta\rho/\rho$ 
grows as $a^{2\ell}$, while in the radiation dominated era 
($a\gg a_{\rm e}$), $\delta\rho/\rho$ is approximately constant. 
The amplification is the same as (\ref{kruno2}).

The evolution of the tensor perturbations for super-horizon modes
is given by equation 
(\ref{bake45}):
\begin{equation}
\tilde{h}=A_{6}+B_{6}\int{(a^{2\ell}+
a_{\rm e}^{2\ell})^{1/\ell}a_{\rm e}\over a^{4}}da
\,.\label{zo4a}\end{equation}
For the inflationary era ($a\ll a_{\rm e}$) equation (\ref{zo4a}) gives 
\begin{equation}
\tilde{h}\approx A_{6}-{1\over 3}B_{6}\left({a_{\rm e}\over a}\right)^{3}
\,,\label{zo5}
\end{equation}
while for the radiation dominated era ($a\gg a_{\rm e}$)
\begin{equation}
\tilde{h}\approx A_{6}-B_{6}\left({a_{\rm e}\over a}\right)\,.\label{zo6}
\end{equation}
Unlike the scalar the tensor perturbations do not grow during  
inflation on super-horizon scales.
\section{Sub-horizon perturbations}
For perturbations on scales inside the Hubble radius (where 
$k/(aH)$ is not negligible), equations (\ref{12e}), (\ref{12a}), 
(\ref{lm1}), (\ref{20}), (\ref{12a1}) and (\ref{23})
can not be solved exactly. We can however give analytic forms for the 
solutions 
in inflation and radiation domination using \cite{M4,M10}, 
since the equations reduce to Bessel forms. 
\subsection{Power-law-like model}
Defining the comoving wave numbers 
$$k_{1}\equiv {a_{\rm e}(1-n)\over t_{\rm e}(1+1/m)^{(1-2n)/2}}\,,$$ 
$$k_{2}\equiv {a_{\rm e}\over 2t_{\rm e}(1+m)^{(1-2n)/2}}\,,$$
we find the following. During power-law-like inflation ($t\ll mt_{\rm e}$), 
the
scalar and tensor perturbations evolve as 
\begin{eqnarray}
\tilde{\Phi} & \approx & \left({t_{\rm e}\over t}\right)^{(n+1)/2}
Z_{(n+1)/2(1-n)}\left[{k\over k_{1}}\left({t\over t_{\rm e}}\right)^{1-n}\right]
\,,\label{mz7}\\
{\delta\rho\over \rho} & \approx & \left[{-2\over 3}\left({k\over k_{1}}
\right)^{2}(1-n)^{2} 
\left({t\over t_{\rm e}}\right)^{(2-2n)}
-2\right]
\left({t_{\rm e}\over t}\right)^{(n+1)/2}Z_{(n+1)/2(1-n)}\left[{k\over k_{1}}
\left({t\over t_{\rm e}}\right)^{(1-n)}\right]
\nonumber\\&&{}+{2(1-n)\over n}{k\over k_{1}}
\left({t\over t_{\rm e}}\right)^{(1-3n)/2}
Z_{(3-n)/2(1-n)}\left[{k\over k_{1}}
\left({t\over t_{\rm e}}\right)^{(1-n)}\right]
\,,\label{mz9}\\
\tilde{h} & \approx & \left({t\over t_{\rm e}}\right)^{(1-3n)/2}
Z_{(3n-1)/2(1-n)}
\left[{k\over k_{1}}\left(
{t\over t_{\rm e}}\right)^{1-n}\right]\,.\label{mz11}\end{eqnarray}
During radiation domination ($t\gg mt_{\rm e}$), we find the following:
\begin{eqnarray}
\tilde{\Phi} & \approx & \left({t_{\rm e}\over t}\right)^{3/4}
Z_{3/2}\left[{k\over k_{2}}\left({t\over t_{\rm e}}\right)^{1/2}\right]
\nonumber\\
&=& \left({t_{\rm e}\over t}\right)^{3/2}\left\{
A_{7}
\sin\left[{k\over k_{2}}\left({t\over t_{\rm e}}\right)^{1/2}
\right]
-B_{7}
\cos\left[{k\over k_{2}}\left({t\over t_{\rm e}}\right)^{1/2}
\right]\right\}\nonumber\\
&&{}-{k\over k_{2}}\left({t_{\rm e}\over t}\right)\left\{A_{7}
\cos\left[
{k\over k_{2}}\left({t\over t_{\rm e}}\right)^{1/2}
\right]+B_{7}
\sin\left[{k\over k_{2}}\left({t\over t_{\rm e}}\right)^{1/2}
\right]\right\}
\,,\label{mz8}\\
{\delta\rho\over \rho} & \approx &
\left[{-8\over 3}\left({k\over k_{\rm 2}}\right)^{2}
{t\over t_{\rm e}}-2\right]
\left({t_{\rm e}\over t}\right)^{3/4}
Z_{3/2}\left[{k\over k_{2}}\left(
{t\over t_{\rm e}}\right)^{1/2}\right]\nonumber\\
&&{}+2{k\over k_{2}}
\left({t_{\rm e}\over t}\right)^{3/4}
Z_{5/2}\left[{k\over k_{2}}\left({t\over t_{\rm e}}\right)^{1/2}
\right]\nonumber\\
& = & 
\left[{-8\over 3}\left({k\over k_{2}}\right)^{2}{t\over t_{\rm e}}-2\right]
\left({t_{\rm e}\over t}\right)^{3/2}\left\{A_{7}\sin
\left[{k\over k_{2}}
\left({t\over t_{\rm e}}\right)^{1/2}\right]
-B_{7}\cos
\left[{k\over k_{2}}
\left({t\over t_{\rm e}}\right)^{1/2}\right]\right\}\nonumber\\
&&{}+\left[{8\over 3}\left({k\over k_{2}}\right)^{2}
{t\over t_{\rm e}}+2\right]
{t_{e}\over t}{k\over k_{2}}\left\{A_{7}\cos
\left[{k\over k_{2}}
\left({t\over t_{\rm e}}\right)^{1/2}\right]
+B_{7}\sin
\left[{k\over k_{2}}
\left({t\over t_{\rm e}}\right)^{1/2}\right]\right\}\nonumber\\
&&{}-{k\over k_{2}}A_{8}\left({t_{\rm e}\over t}\right)^{2}
\left\{3\sin\left[{k\over k_{2}}
\left({t\over t_{\rm e}}\right)^{1/2}\right]
-3{k\over k_{2}}
\left({t\over t_{\rm e}}\right)^{1/2}\cos
\left[{k\over k_{2}}
\left({t\over t_{e}}\right)^{1/2}\right]\right.\nonumber\\ 
&&{}-\left.
\left({k\over k_{2}}\right)^{2}{t\over t_{\rm e}}\sin
\left[{k\over k_{2}}
\left({t\over t_{\rm e}}\right)^{1/2}\right]\right\}\nonumber\\
&&{}-{k\over k_{2}}B_{8}\left({t_{\rm e}\over t}\right)^{2}\left\{3\cos
\left[{k\over k_{2}}\left({t\over t_{\rm e}}\right)^{1/2}\right]
+3{k\over k_{2}}
\left({t\over t_{\rm e}}\right)^{1/2}\sin
\left[{k\over k_{2}}
\left({t\over t_{\rm e}}\right)^{1/2}\right]\right.\nonumber\\ &&{}-
\left. 
\left({k\over k_{2}}\right)^{2}{t\over t_{e}}\cos
\left[{k\over k_{2}}
\left({t\over t_{\rm e}}\right)^{1/2}\right]\right\}
\,,\label{mz10}\\
\tilde{h} & \approx &
\left({t_{\rm e}\over t}\right)^{1/4}Z_{1/2}\left[{k\over k_{2}}\left({
t\over t_{\rm e}}\right)^{1/2}\right]\nonumber\\
& = & \left({t_{\rm e}\over t}\right)^{1/2}
\left\{A_{9}\sin\left[{k\over k_{2}}
\left({t\over t_{\rm e}}\right)^{1/2}\right]
-B_{9}\cos\left[{k\over k_{2}}
\left({t\over t_{e}}\right)^{1/2}\right]\right\}\,.
\label{mz12}\end{eqnarray}
As expected, density perturbations oscillate after re-entry, since radiation 
pressure balances gravitational infall. Note that while there are non-decaying 
oscillations in the density perturbations, the tensor perturbation 
oscillations are purely decaying. At late times, the density perturbation 
(\ref{mz10}) has the form 
$${\delta{\rho}\over \rho}\approx 
A\cos\left({k\over k_{2}}{a\over a_{\rm e}}\right)
+B\sin\left({k\over k_{2}}
{a\over a_{\rm e}}\right)\,,$$
in agreement with the standard results for the radiation era \cite{nh1}
(p. 152).
\subsection{de Sitter-like model}
Sub-horizon perturbations during inflation ($a\ll a_{\rm e}$) have the form
\begin{eqnarray} 
\tilde{\Phi} & \approx & \left({a\over a_{\rm e}}
\right)^{(2\ell-1)/2}
Z_{-(2\ell+1)/2}\left(-{{\pi\over 2^{(1-\ell)/\ell}}{k\over k_{\rm e}}{a_{e}\over a}}\right)
\,,\label{lm1a}\\
{\delta\rho\over \rho} & \approx & -{2\over 3}
\left({{\pi\over 2^{(1-\ell)/\ell}}
{k\over k_{\rm e}}}
\right)^{2}
\left({a\over a_{\rm e}}\right)^{(2\ell-5)/2}
Z_{-(2\ell+1)/2}\left({-{\pi\over 2^{(1-\ell)/\ell}}
{k\over k_{\rm e}}{a_{\rm e}\over a}}\right)\nonumber\\
&&{}-
{2\pi\over 2^{(1-\ell)/\ell}}{k\over k_{\rm e}}
\left({a\over a_{\rm e}}\right)^{(2\ell-3)/2}
Z_{-(2\ell+3)/2}\left({-{\pi\over 2^{(1-\ell)/\ell}}
{k\over k_{\rm e}}{a_{\rm e}\over a}}\right)
\,,\label{lm3}\\
\tilde{h} & \approx & \left({a_{\rm e}\over a}\right)^{3/2}
Z_{-3/2}\left(-{\pi\over 2^{(1-\ell)/\ell}}{k\over k_{\rm e}}
{a_{\rm e}\over a}\right)
\nonumber\\ & = & A_{10}\left[\cos\left({\pi\over 2^{(1-\ell)/\ell}}
{k\over k_{\rm e}}{a_{\rm e}\over a}\right)+{\pi\over 2^{(1-\ell)/\ell}}
{k\over k_{\rm e}}{a_{\rm e}\over a}\sin\left({\pi\over 2^{(1-\ell)/\ell}}
{k\over k_{\rm e}}{a_{\rm e}\over a}\right)\right]\nonumber\\
&&{}-B_{10}\left[\sin\left({\pi\over 2^{(1-\ell)/\ell}}
{k\over k_{\rm e}}{a_{\rm e}\over a}\right)+{\pi\over 2^{(1-\ell)/\ell}}
{k\over k_{\rm e}}{a_{\rm e}\over a}\sin\left({\pi\over 2^{(1-\ell)/\ell}}
{k\over k_{\rm e}}{a_{\rm e}\over a}\right)\right]\,.\label{lm5}
\end{eqnarray}
Sub-horizon perturbations during radiation domination 
($a\gg a_{\rm e}$) are given by:
\begin{eqnarray}
\tilde{\Phi} & \approx & \left({a_{\rm e}\over a}\right)^{3/2}
Z_{3/2}\left({\pi\over 2^{(1-\ell)/\ell}}{k\over k_{\rm e}}{a\over a_{\rm e}}
\right)
\nonumber\\
& = & A_{11}\left({a_{\rm e}\over a}\right)^{3}\left[\sin\left({\pi\over 
2^{(1-\ell)/\ell}}
{k\over k_{\rm e}}{a\over a_{\rm e}}
\right)-
{\pi\over 2^{(1-\ell)/\ell}} {k\over k_{\rm e}}{a\over a_{\rm e}}
\cos\left({\pi\over 2^{(1-\ell)/\ell}}{k\over k_{\rm}}
{a\over a_{\rm e}}\right)\right]\nonumber\\
&&{}-B_{11}\left({a_{\rm e}\over a}\right)^{3}\left[
\cos\left({\pi\over 2^{(1-\ell)/\ell}}{k\over k_{\rm e}}{a\over a_{\rm e}}
\right)+
{\pi\over 2^{(1-\ell)/\ell}}{k\over k_{\rm e}}
{a\over a_{\rm e}}
\sin\left({\pi\over 2^{(1-\ell)/\ell}}{k\over k_{\rm e}}{a\over a_{\rm e}}\right)
\right]
\,,\label{lm2}\\
{\delta\rho\over \rho} & \approx & \left[-{2\over 3}
\left({\pi\over 2^{(1-\ell)/\ell}}{k\over k_{\rm e}}\right)^{2}
\left({a\over a_{\rm e}}\right)^{2}
-2\right]\left({a_{\rm e}\over a}\right)^{3/2}
Z_{3/2}\left({\pi\over 2^{(1-\ell)/\ell}}
{k\over k_{\rm e}}{a\over a_{\rm e}}
\right)
\nonumber\\ &&{}+
{{2\pi\over 2^{(1-\ell)/\ell}}{k\over k_{\rm e}}}
\left({a_{\rm e}\over a}\right)
^{1/2}Z_{5/2}\left({\pi\over 2^{(1-\ell)/\ell}}{k\over k_{\rm e}}
{a\over a_{\rm e}}\right)
\nonumber\\
& = & \left[{-2\over 3}\left({\pi\over 2^{(1-\ell)/\ell}}{k\over k_{\rm e}}
\right)^{2}
\left({a\over a_{\rm e}}\right)^{2}
-2\right]\left({a_{\rm e}\over a}
\right)^{3}A_{11}
\left[\sin\left({\pi\over 2^{(1-\ell)/\ell}}{k\over k_{\rm e}}
{a\over a_{\rm e}}\right)\right.
\nonumber\\ &&{}-
\left.{\pi\over 2^{(1-\ell)/\ell}}{k\over k_{\rm e}}{a\over a_{\rm e}}
\cos\left({\pi\over 2^{(1-\ell)/\ell}}{k\over k_{\rm e}}{a\over a_{\rm e}}\right)
\right]\nonumber\\
&&{}+\left[{2\over 3}\left({\pi\over 2^{(1-\ell)/\ell}}
{k\over k_{\rm e}}\right)^{2}
\left({a\over a_{\rm e}}\right)^{2}+2\right]
\left({a_{\rm e}\over a}\right)^{3}
B_{11}\left[
\cos\left
({\pi\over 2^{(1-\ell)/\ell}}{k\over k_{\rm e}}{a\over a_{\rm e}}\right)
\right.
\nonumber\\ &&{}+ 
\left.
{\pi\over 2^{(1-\ell)/\ell}}{k\over k_{\rm e}}{a\over a_{\rm e}}\sin\left
({\pi\over 2^{(1-\ell)/\ell}}{k\over k_{\rm e}}{a\over a_{\rm e}}\right)
\right]
\nonumber\\
&&{}-A_{12}\left({a_{\rm e}\over a}\right)^{3}
\left[\left({\pi\over 2^{(1-\ell)/\ell}}{k\over k_{\rm e}}{a\over a_{\rm e}}
\right)^{2}
\sin\left({\pi\over 2^{(1-\ell)/\ell}}{k\over k_{\rm e}}{a\over a_{\rm e}}
\right)
\right.\nonumber\\&&{}-\left. 3\sin\left({\pi\over 2^{(1-\ell)/\ell}}
{k\over k_{\rm e}}{a\over a_{\rm e}}\right)
\right.\nonumber\\ &&{}+
\left.
3{{\pi\over 2^{(1-\ell)/\ell}}{k\over k_{\rm e}}{a\over a_{\rm e}}}
\cos\left({\pi\over 2^{(1-\ell)/\ell}}{k\over k_{\rm e}}
{a\over a_{\rm e}}\right)\right]\nonumber\\
&&{}+B_{12}\left({a_{\rm e}\over a}\right)^{3}\left
[\left({\pi\over 2^{(1-\ell)/\ell}}
{k\over k_{\rm e}}{a\over a_{\rm e}}\right)^{2}\cos\left
({\pi\over 2^{(1-\ell)/\ell}}{k\over k_{\rm e}}{a\over a_{\rm e}}\right)
\right.\nonumber\\&&{}-\left. 3\cos\left({\pi\over 2^{(1-\ell)/\ell}}
{k\over k_{\rm e}}{a\over 
a_{\rm e}}\right)-
{3\pi\over 2^{(1-\ell)/\ell}}{k\over k_{\rm e}}
{a\over a_{\rm e}}
\sin\left
({\pi\over 2^{(1-\ell)/\ell}}{k\over k_{\rm e}}{a\over a_{\rm e}}\right)
\right]
\,,\label{lm4}\\
\tilde{h} & \approx & \left({a_{\rm e}\over a}\right)^{1/2}
Z_{1/2}\left({\pi\over
2^{(1-\ell)/\ell}}{k\over k_{\rm e}}{a\over a_{\rm e}}\right)
\nonumber\\ & = & {a_{\rm e}\over a}
\left[A_{13}\sin\left({\pi\over 2^{(1-\ell)/\ell}}{k\over k_{\rm e}}
{a\over a_{\rm e}}\right)
-B_{13}\cos\left({\pi\over 2^{(1-\ell)/\ell}}{k\over k_{\rm e}}{a\over a_{\rm e}}\right)\right]
\,.\label{lm6}\end{eqnarray}
Once again, we see that the density perturbations contain non-decaying
oscillations while the tensor perturbation oscillations are purely 
decaying. At late times, the density perturbation (\ref{lm4}) has the 
form
$${\delta{\rho}\over \rho}\approx A\cos
\left({\pi\over 2^{(1-\ell)/\ell}}
{k\over k_{\rm e}}{a\over a_{\rm e}}\right)+
B\sin
\left({\pi\over 2^{(1-\ell)/\ell}}
{k\over k_{\rm e}}{a\over a_{\rm e}}\right)\,,$$
as in the power-law-like case.
\section{Conclusions}
Our main result is a comprehensive presentation of analytic solutions for the scalar 
and tensor perturbations on super- 
and sub-horizon scales 
for two classes of cosmology, where the 
scale factor evolves smoothly from an inflationary era to a 
radiation dominated era. These classes encompass parametrised
power-law-like and de Sitter-like inflationary behaviour.
Our 
results confirm the amplification of super-horizon scalar perturbations 
for all values of the parameters which define the classes of models.
Explicit forms, given in equations (\ref{kruno}) and (\ref{kruno2}), 
of the amplification for scalar perturbations show how 
the de Sitter-like inflation is much more effective in amplification than 
the power-law-like inflation. On sub-horizon scales, the detailed form of
oscillating modes after re-entry was found, showing that there are 
non-decaying
modes for density perturbations, but not for tensor perturbations.
Taken together, 
these results provide a unified catalogue of perturbation solutions 
in what can be taken as the two main types of inflationary models
with smooth exit.

\end{document}